# PROPOSTA DE UM SISTEMA ELETRÔNICO DE AUDITORIA APLICADO À URNA ELETRÔNICA BRASILEIRA

**Guimarães, Marcelo Ferreira**
Fundação CERTI – Campus da UFSC
Caixa Postal 5053 – Florianópolis – SC
mfg@certi.ufsc.br

**Sell, Carlos Antônio**
Fundação CERTI – Campus da UFSC
Caixa Postal 5053 – Florianópolis – SC
cae@certi.ufsc.br

**Turcato, Renato Parenti**
Fundação CERTI – Campus da UFSC
Caixa Postal 5053 – Florianópolis – SC
rpt@certi.ufsc.br

**Assuiti, Carlos Henrique**
Fundação CERTI – Campus da UFSC
Caixa Postal 5053 – Florianópolis – SC
cha@certi.ufsc.br

**Custódio, Ricardo Felipe**
Laboratório de Segurança em Computação – LabSEC
Departamento Informática e Estatística da UFSC
Campus da UFSC – Florianópolis – SC
custodio@inf.ufsc.br

**Santos, Ricardo Antônio Pralon**
Fundação CERTI – Campus da UFSC
Caixa Postal 5053 – Florianópolis – SC
rap@certi.ufsc.br

## RESUMO

*Um novo sistema, chamado SELA - Sistema Eletrônico de Auditoria foi desenvolvido para ser aplicado nas Urnas Eletrônicas Brasileiras. O SELA foi concebido como um sistema aberto, tanto o Hardware como o Software, para que seja amplamente conhecido e divulgado pela Sociedade. A segurança do Processo de Auditoria dos dados captados da Urna Eletrônica é garantida através do uso de algoritmos consagrados que implementam uma Função Resumo (Hash). Este artigo descreve o SELA bem como analisa a sua aplicação durante o processo eleitoral. Uma análise comparativa é feita entre o SELA e Impressoras Térmicas como sistema secundário de registro dos votos. Os autores recomendam a aplicação piloto do SELA nas Eleições de 2002.*

## ABSTRACT

*A new system, called SELA - Auditing Electronic System, has been developed in order to be applied to the Brazilian Electronic Voting Machine. The SELA was conceived to use an open Hardware and Software to be well known by the Society. The security of the auditing process is guaranteed by the application of a Fingerprint Algorithm, a Hash Function. The system is very robust and requires minimum modification of the Electronic Voting Machine. In this paper, the SELA is described as well as its use during the election process. A comparison between SELA and the use of thermal printers has also been made. A pilot application of SELA for the 2002 Elections is recommended by the authors.*

## 1 INTRODUÇÃO

Nos últimos 10 anos, a Sociedade Brasileira fez investimentos significativos na informatização do Processo Eleitoral Brasileiro, com ganhos visíveis de eficiência e eficácia.

Hoje o país é reconhecido internacionalmente como um líder na área de Automação Eleitoral, pois conseguiu realizar em 2000 a maior eleição informatizada do mundo, quando aproximadamente mais de 108 milhões eleitores participaram do processo e, em poucas horas, os resultados da eleição foram conhecidos.

A Urna Eletrônica, - UE que é para o cidadão a parte mais visível de todo o processo eleitoral informatizado, contribuiu para esse sucesso por se tratar de um sistema robusto e muito simples de ser usado. É um exemplo prático de equipamento que promoveu a inclusão digital no país, sendo utilizado por cidadãos de todas as regiões, classes sociais e níveis de escolaridade.

A facilidade de absorção de novas tecnologias pela Sociedade Brasileira é hoje apontada por vários estudos de classificação de grau de desenvolvimento entre países (estudo do IMD em referência) como um fator de competitividade do Brasil. Exemplos como da Urna Eletrônica, Declarações de Imposto de Renda e Popularização de Terminais de Auto-atendimento Bancário são freqüentemente citados na mídia como casos de sucesso e confirmam que a População Brasileira não é avessa a Inovações Tecnológicas.

Especificamente no caso da Votação Eletrônica, o país assumiu um papel de líder inovador mundial e como tal tem a responsabilidade de manter o ritmo de evolução tecnológica das soluções adotadas.

Após a consolidação do uso da Urna Eletrônica como um sistema de captação da vontade popular em 2000, pouco se tem discutido de forma sistemática pela comunidade técnico-científica em relação ao futuro tecnológico do Sistema de Automação das Eleições Brasileiras.

Após o caso de quebra de sigilo do Painel Eletrônico do Senado, iniciou-se uma discussão na mídia se existiriam problemas de segurança também na urna. Os poucos Fóruns de Discussão no Brasil sobre os melhoramentos e evolução tecnológica das Urnas Eletrônicas foram muito politizados, prejudicando a qualidade técnica da discussão.

É fato que a Sociedade Brasileira hoje exige que a comunidade técnico-científica ajude a esclarecer se o sistema adotado pelo país é seguro, pois este agora é o tema central nas próximas eleições.

Já não é mais necessário discutir se a Urna vai ser útil ou não, ou se os cidadãos vão conseguir votar ou não, ou se ela vai funcionar ou não, como eram as preocupações nas eleições de 1996, 1998 e 2000.

Essas perguntas já estão respondidas pelo sucesso da aplicação das Urnas. A questão agora é saber se as Urnas são realmente seguras e como garantir ao cidadão o direito de verificar que seu voto será contabilizado de forma correta.

Este trabalho tem o objetivo de propor uma solução inovadora para mostrar que a urna é segura. A Urna Eletrônica tem que ser imune a ataques externos e a ataques internos.

Neste sentido, foi desenvolvido o SELA - Sistema Eletrônico para Auditoria - que aplicado à Urna Eletrônica resulta num sistema que atende um maior número de requisitos de segurança. O desenvolvimento do SELA deu atenção especial aos aspectos relativos à transparência do processo eleitoral. O sistema utiliza o *hardware* existente, muda o mínimo do *software* e evita os problemas do papel.

O trabalho apresenta a solução enfatizando seus conceitos e características de segurança. O SELA é visto como uma solução de curto prazo para alguns questionamentos importantes feitos pela sociedade, como a garantia de auditoria externa ao sistema no dia da votação e permitir a visualização do voto pelo eleitor.

## 2 SISTEMA ELETRÔNICO DE AUDITORIA – SELA: UMA ALTERNATIVA INOVADORA

O SELA foi desenvolvido a partir de uma analogia feita com um acessório, atualmente existente na urna eletrônica, denominado Urna Plástica Descartável – UPD.

### *2.1 Urna Plástica Descartável – UPD*

A UPD (figura 1) é constituída por um saco plástico preto para armazenar os votos impressos, sustentado por uma estrutura em poliestireno com bocais de encaixe e fixação projetados para conexão mecânica com a Urna Eletrônica. Cumpre funções similares às da antiga urna de lona e caracteriza-se por:

- Permitir a conferência dos resultados da UE;
- Ser um dispositivo para depósito/armazenagem de forma segura dos votos impressos em papel pela UE;
- Permitir a constatação visual de ZERO votos depositados no início da votação;
- Possibilitar seu uso após o encerramento da votação com identificação da seção;
- Ter lacre e procedimentos jurídicos que garantem a segurança dos votos fisicamente armazenados;
- Ser conectável a qualquer UE do mesmo modelo;
- Ter baixo custo.

Estas características permitem que o conjunto UE/UPD confira maior transparência ao processo técnico de votação e credibilidade junto ao eleitor. Isto acontece porque para o eleitor a UPD constitui-se num sistema paralelo, simples e conhecido de conferência e, assim, o funcionamento interno da UE passa a ser secundário.

No entanto, alguns fatores criam limitações significativas para esta solução. São eles:

- Existem sérios problemas técnicos em relação ao transporte, corte e armazenagem de papel que diminuem a robustez do sistema como um todo;
- O uso da UPD no processo eleitoral permite a volta de ataques e fraudes clássicos na era da urna de lona;
- A visualização do voto impresso antes do depósito não é possível na UPD atualmente implementada.
- O voto não é assinado pela mesa não tendo valor jurídico.

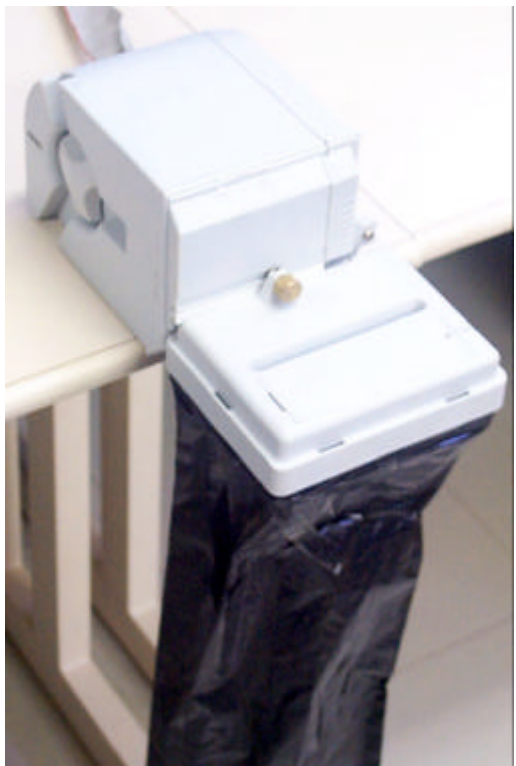

Figura 1 – Foto da UPD.

Tudo isto foi considerado para a não utilização da UPD e para desenvolver-se o SELA como uma alternativa.

### 2.2 Solução Conceitual

O SELA é um dispositivo eletrônico que tem características similares a de uma UPD. Conceitos de segurança tais como: função resumo e arquitetura aberta; e conceitos de projeto robusto tais como: poucas peças, sem partes móveis, uso de tecnologias dominadas e foco nas interfaces foram empregados neste dispositivo.

#### 2.2.1 Requisitos Adotados

No quadro 1 são apresentados os principais requisitos empregados para o desenvolvimento do SELA.

#### 2.2.2 Concepção Física do Dispositivo

O SELA (figura 2) é composto por uma placa eletrônica e um visor acomodados num gabinete de plástico e conectado a UE por uma interface. Seu "design" deve ser ergonômico de modo que o eleitor possa visualizar e comparar claramente os números mostrados no visor.

**Quadro 1: Principais requisitos do SELA**

| | |
|---|---|
| **Arquitetura Aberta** | Projeto de *Hardware* e *Software* aberto ao conhecimento público. |
| **Uso de Primitivas Criptográficas** | Segurança baseada em algoritmos criptográficos de conhecimento público e padrões de segurança internacionais. |
| **Auditável** | Projetado de forma a ser facilmente auditado por parte dos partidos, Tribunal Superior Eleitoral - TSE e sociedade a qualquer momento. |
| **Independente de eleição e de urna** | Não necessita de qualquer programação antes da eleição ou após a eleição. |
| **Mínimo de modificações no sistema atual da urna** | Permitir o uso na maioria das urnas já fabricadas de forma a preservar o patrimônio público. |
| **Logística simples** | Minimizar as alterações necessárias aos processos de logística usados atualmente. |
| **Uso em 100% e/ou por amostragem** | Auditoria em amostragem ou 100% conforme grau de segurança desejado. |
| **Baixo custo** | Dispositivo de baixo custo (cerca de 10% do custo da Urna Eletrônica). |
| **Visualização do Voto** | Permitir que o eleitor veja ergonomicamente o voto digitado comparando-o com o mostrado na UE. |

O gabinete deve ser transparente permitindo a inspeção visual dos componentes eletrônicos, seu sistema de fechamento e conexões de interface devem ser projetadas para permitirem o uso de lacres.

#### 2.2.3 Arquitetura de "Hardware" e Descrição dos Estados

Na figura 3 são visualizados os seguintes módulos:

- CPU - Unidade de Processamento - tem a função de executar o código gravado;
- MEMÓRIA - armazena o código do programa, que pode ser facilmente lido com instrumentos existentes no mercado. Acumula os dados dos totalizadores dos votos de forma não seqüencial;
- VISOR - Mostrador com capacidade de apresentar simultaneamente 5 números decimais, sendo assim capaz de mostrar todos os números de candidatos/legenda existentes no país;
- INTERFACE de ENTRADA/SAÍDA - Interface elétrica com a UE, executa a comunicação da UE com o SELA. Tem Interface onde são extraídos os totalizadores de votos em caso de auditoria.

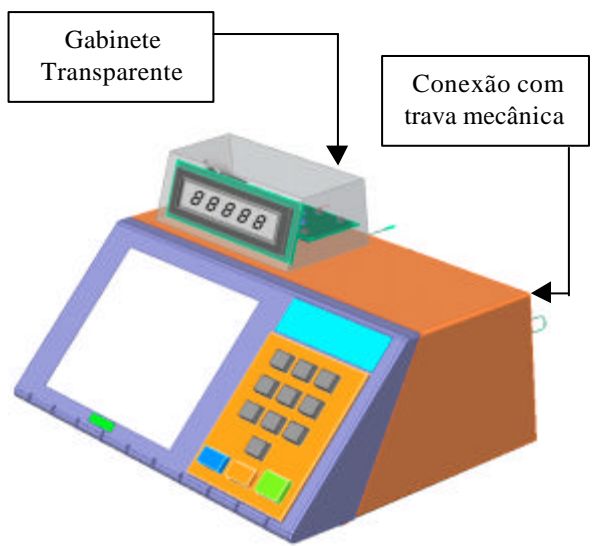

Figura 2 – Concepção Física do SELA

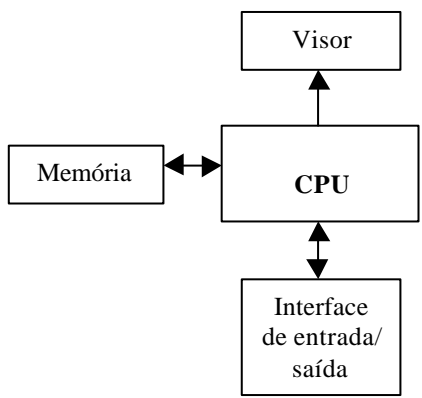

Figura 3 - Arquitetura básica do *Hardware* Eletrônico do SELA

O funcionamento do SELA é descrito a seguir:

- Inicialização: a UE envia comando de início para o SELA, toda a memória é limpa e as variáveis inicializadas;
- Zerésima: a UE envia comando para cálculo da função resumo do *software* do SELA. A função resumo é calculada e mostrada no visor para ser anotada pelos mesários;
- Votação: A UE envia comando de abertura de eleitor marcando início de um voto. A UE envia comando de votação com a identificação do pleito e número do candidato a ser mostrado no visor;
- número do candidato é armazenado em memória e pode:
  - ser CORRIGIDO : apaga número do visor e da memória;
  - ser CONFIRMADO: armazena número na memória e incrementa quantidade de votos para o candidato;
- A UE envia comando para fechar o voto do eleitor;
- Finalização: A UE envia comando para término da eleição. O SELA calcula a função resumo dos votos contidos na memória e mostra esse valor no visor para anotação por parte dos mesários. O SELA entra em modo de Auditoria;
- Auditoria: O SELA apenas permite a leitura dos votos (pleito, número do candidato e número de votos) contidos na memória. Os dados são gravados na memória. O valor da função resumo calculado é comparado ao valor anotado pelos mesários no fechamento da votação e garante a legitimidade do SELA.

### 2.2.4 O Conceito de Segurança

Para garantir que os dados do SELA sejam os mesmos publicados e, portanto, não sejam maliciosamente alterados, é necessário estabelecer um mecanismo que possibilite a obtenção de uma "assinatura" de seu código e dados. Esta assinatura é possível através da utilização de uma função *hash* ou função resumo. Funções resumo, permitem que, ao ser aplicada à uma mensagem de qualquer tamanho, seja gerado um resumo de tamanho fixo e bastante pequeno (normalmente 128 ou 160 *bits*).

Com a função resumo pode-se garantir que o conteúdo de uma mensagem não foi alterado. Ela representa um papel fundamental na criptografia moderna pois permite garantir a integridade de dados e a autenticação de mensagens.

A idéia básica das funções resumo é que um resumo serve como uma imagem representativa compacta (às vezes chamada de impressão digital ou "*message digest*") da cadeia de *bits* da entrada, e pode ser usada como se fosse unicamente identificável com aquela entrada. As funções resumo funcionam semelhante ao dígito verificador do CPF. Por exemplo, se um número qualquer do CPF for modificado, o dígito verificador também sofrerá alteração.

Uma boa função resumo tem uma característica chamada de efeito avalanche. Isto significa que uma pequena mudança no arquivo de entrada acarreta uma grande e imprevisível mudança na saída. As funções resumo são diferentes das funções normais de criptografia por não possuírem uma chave e por serem irreversíveis. Qualquer um pode verificar a autenticidade de uma certa mensagem, apenas calculando a função novamente, e comparando o resultado com o já obtido anteriormente.

Assim, as funções de resumo podem ser usadas para garantir a integridade do código e dos dados do SELA. A função resumo deve ser conhecida de todos e aceita sem restrições quanto a sua funcionalidade.

Por isto optou-se pelo uso da função SHA-1. Esta função é aceita internacionalmente como segura, eficaz e à prova de fraudes.

O procedimento de uso da função resumo no SELA deve ser o seguinte:

1. Os mesários, juntamente com os fiscais dos partidos políticos, devem calcular através do SELA, o resumo do código e dados do SELA. Este valor já é previamente conhecido e deve coincidir com o valor previamente publicado e, portanto, de conhecimento público. Esse procedimento irá garantir que o código é o original e que não há dados previamente inseridos no SELA. Assim, garante-se a independência do SELA em relação a URNA e a uma Eleição em particular;
2. Ao final da votação, novamente é aplicada a função resumo. O resultado desta função também é anotado pela mesa e pelos fiscais dos partidos. O valor deve ser anotado em documento formal e assinado por todos os presentes. Isso irá garantir que a partir deste momento não será mais possível alterar qualquer informação contida no SELA.

### 2.3 O Protótipo Implementado

Para testar o conceito proposto foi implementado um protótipo do SELA, que é mostrado na figura 4.

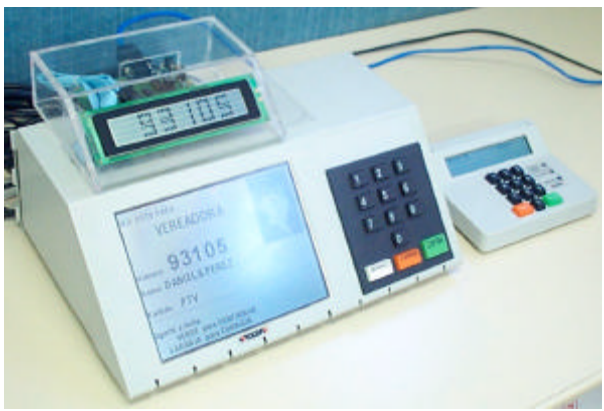

Figura 4 - Foto do protótipo implementado para testar o conceito

Foram utilizados componentes comerciais disponíveis no mercado e desenvolvido um programa para simular a eleição em um microcomputador. Foi utilizada uma interface serial padrão RS232 para comunicação entre o computador e o SELA.

### *2.4 O Processo de Uso*

O SELA foi desenvolvido para ser utilizado durante a eleição na seção eleitoral e, posteriormente, no processo de auditoria da urna.

### *2.4.1 O Uso do SELA na Seção Eleitoral*

O quadro 2 mostra as etapas de utilização do SELA no dia da votação em uma seção eleitoral.

**Quadro 2: Uso do SELA na seção eleitoral.**

| Processo | Quem? | Urna | SELA |
|---|---|---|---|
| Instalação | Mesários | Conectar a energia elétrica. | Conectar com a Urna e lacre da interface. |
| | | Autoteste. | Autoteste. |
| | | Verificação dos lacres. | "Zeragem" da memória de dados. |
| Zerésima | Mesários | Impressão do relatório informando que todos os candidatos tem ZERO votos. | Mesários devem anotar na ATA da Eleição o resultado do cálculo da Função Resumo do código de programa mais a memória de dados. Este número será o mesmo para todo Brasil. |
| Início de Votação | Automático as 8 horas. | Envia para o SELA as informações da seção. | Grava as informações da Seção. |
| | | Disponibiliza a entrada do número do título de eleitor no microterminal. | Aguarda comando da urna informando inicio de votação do eleitor. |
| Votação | Mesário | Habilita a votação de um novo eleitor. | Recebe comando da Urna informando início do processo de votação de novo eleitor. |
| | Eleitor | Tecla o número do candidato escolhido para o pleito. | Recebe da urna os números teclados pelo eleitor para cada pleito. |
| | | | O eleitor confere se o número que aparece na urna é o mesmo que aparece no visor do SELA podendo confirmar ou corrigir seu voto. O SELA grava em sua memória de dados o último número mostrado no visor (voto). |
| Finalização do Pleito | Mesário | Emite o Boletim de Urna com os resultados da seção. | Finaliza o processo com o cálculo da função resumo dos dados da memória. |
| | | Informa ao SELA o final de eleição. | Mesário anota na ATA da eleição o resultado da função resumo calculada. |
| | | Grava disquete com o resultado. | Convoca fiscais para verificarem o resultado da função resumo e assinam ATA da eleição. |
| Transporte do Disquete e Embalagem | Mesário | Levar disquete para local de apuração. | Desconecta SELA, embala e entrega junto com o disquete. |

### 2.4.2 O Uso do SELA para Auditoria das Urnas Eletrônicas

Durante a auditoria de uma seção deve-se:

- Verificar a integridade física do SELA através dos lacres;
- Verificar anotações do resultado da função resumo feitos pelos mesários da seção e fiscalizados pelos partidos;
- Ler o código do programa do microcontrolador do SELA e compará-lo com o do código fonte aberto;
- Extrair dados da votação do SELA utilizando um comando específico para isto;
- Comparar dados finais do SELA com os dados da urna da seção;
- Facultativamente pode-se recalcular "manualmente" a função resumo da Zerésima;
- Facultativamente pode-se recalcular "manualmente" a função resumo do resultado da eleição na seção.

### 2.5 Análise de Robustez da Solução Conceitual

O quadro 3 demonstra a robustez do SELA em relação a algumas das principais perturbações às quais estará exposto durante seu ciclo de vida.

### 2.6 Análise Comparativa: Voto Impresso x SELA

O quadro 4 apresenta uma comparação entre o uso do voto impresso e o SELA como mecanismos de auditoria e conferência da votação.

**Quadro 3: Análise Simplificada de Robustez do SELA**

| Ambiente | Ruídos | Ações |
|---|---|---|
| Fabricação | *Hardware* Adulterado | Homologação de fornecedores |
| | *Firmware* Adulterado | Acompanhamento da produção |
| | Memória de dados pré-gravada | Lacres |
| Transporte | *Hardware* Adulterado | Inspeção de recebimento nos TRE's |
| | *Firmware* Adulterado | Inspeção de recebimento nos TRE's |
| TRE´s e Partidos | *Hardware* Adulterado | Testes 100% com acompanhamento dos partidos |
| | | Inspeção visual do gabinete transparente do SELA |
| | *Firmware* Adulterado | Verificação do CRC do código gravado |
| | | Cálculo da função resumo do código e memória de dados |
| | | Lacres de inviolabilidade |
| Seções Eleitorais | SELA com lacres rompidos | Substituição do SELA |
| | SELA não é conectado à Urna | Seção deixaria de ser auditada |
| | Urna apresenta problema durante votação | Seção deixaria de ser auditada, pois o SELA necessita passar por todas as fases de votação para ter dados válidos. |
| | SELA desconectado durante votação | Urna dá alarme no próximo evento da seção |
| | SELA apresenta problema durante votação | Auditoria cancelada com registro em ATA |
| Zonas de Apuração | Lacres violados | Violação dos lacres deve ser tratado com procedimento jurídico a ser definido |
| | Não é possível ler os dados do SELA (probabilidade de falha é muito pequena) | Seção deixaria de ser auditada |
| | Dados do SELA diferentes da Urna | Incoerência entre dados deve ser tratado com procedimento jurídico a ser definido. Se tudo estiver correto com o SELA é a indicação de problemas com a Urna |

**Quadro 4: Análise Comparativa Voto Impresso x SELA**

| Requisito | Impressão Física do Voto no Papel | SELA |
|---|---|---|
| **Credibilidade e segurança** | - Possibilidade de ataque aos votos depositados durante recontagem;<br>- Possibilidade de adulteração dos votos no transporte. | - *Hardware* e *Software* abertos com segurança garantida por algoritmos matemáticos consagrados;<br>- Sigilo do eleitor garantido por uso de acumuladores. |
| **Desempenho** | - Aumento no tempo de votação devido ao tempo de impressão e verificação;<br>- Correção de voto exige cancelamento de voto impresso ou uso de outra impressão. | - Baixa modificação da urna sem alteração de desempenho e tempo de votação;<br>- Fácil correção do voto. |
| **Robustez** | - Mecanismo sensível com possível enrosco de papel;<br>- Alto consumo de energia da bateria da urna em locais sem energia;<br>- Uso de papel térmico que é sensível às condições de umidade. | - Sistema compacto e sem partes móveis;<br>- Baixíssimo consumo de energia, inferior a 5% do consumo de uma impressora térmica com corte de papel no mesmo ciclo de trabalho;<br>- Dados na forma de *bits*. |
| **Facilidade de uso e operação** | - Baixa ergonomia para leitura do voto (impressão com letras pequenas comparado ao SELA);<br>- Logística e instalação de mais uma impressora e rolo de papel;<br>- Processo complexo de auditoria. | - Ótima ergonomia de conferência pelo eleitor;<br>- Fácil instalação;<br>- Processo muito simples de auditoria. |
| **Custo** | - Custo de logística dos consumíveis;<br>- Custo de impressora e consumível. | - Custo compatível com impressora sem consumível. |

## 3. CAMINHOS FUTUROS E CONCLUSÕES

Recomenda-se que o SELA possa ser implementado já nas próximas eleições de 2002, de forma paulatina, através de um lote piloto.

A partir da avaliação deste programa piloto será possível fazer otimizações e definir os processos de auditoria para as próximas eleições de forma abrangente. Desta forma, os partidos políticos e a sociedade terão uma ferramenta poderosa que garantirá a transparência do processo de apuração das eleições. Considerando a segurança do processo eleitoral, o SELA evita o retorno das fraudes do passado.

A aplicação deste sistema não implica que melhoramentos de segurança da própria urna deixem de ser perseguidos, como a implantação de assinaturas digitais nos *softwares* da urna.

Certamente o SELA contribui para a manutenção da credibilidade da Urna Eletrônica, conquista e patrimônio da Sociedade Brasileira.

## AGRADECIMENTOS